# Joint Segmentation and Image Reconstruction with Error Prediction in Photoacoustic Imaging using Deep Learning

Ruibo Shang[1], Geoffrey P. Luke[2], Matthew O'Donnell[1,*]

[1]uWAMIT Center, Department of Bioengineering, University of Washington, Seattle, WA 98195, USA
[2]Thayer School of Engineering, Dartmouth College, Hanover, NH 03755, USA

*odonnel@uw.edu

## Abstract

Deep learning has been used to improve photoacoustic (PA) image reconstruction. One major challenge is that errors cannot be quantified to validate predictions when ground truth is unknown. Validation is key to quantitative applications, especially using limited-bandwidth ultrasonic linear detector arrays. Here, we propose a hybrid Bayesian convolutional neural network (Hybrid-BCNN) to jointly predict PA image and segmentation with error (uncertainty) predictions. Each output pixel represents a probability distribution where error can be quantified. The Hybrid-BCNN was trained with simulated PA data and applied to both simulations and experiments. Due to the sparsity of PA images, segmentation focuses Hybrid-BCNN on minimizing the loss function in regions with PA signals for better predictions. The results show that accurate PA segmentations and images are obtained, and error predictions are highly statistically correlated to actual errors. To leverage error predictions, confidence processing created PA images above a specific confidence level.

# 1. Introduction

A key step in quantitative photoacoustics (PA) is accurately reconstructing the initial pressure distribution throughout the imaging volume from data collected on an ultrasound array transducer [1, 2]. With this initial pressure distribution (IPD) and the estimated fluence at each pixel, the optical absorption coefficient can be imaged [3-5]. Optical absorption images at different wavelengths can then be combined for spectroscopic imaging, bringing molecular profiling to deep structures within the body [6-8]. By combining quantitative spectroscopic PA imaging with real-time ultrasound (US) in a handheld probe, true molecular sensitivity can be added to clinical US (i.e., real-time PAUS imaging) [9]. Therefore, improving IPD reconstruction is key to quantitative PA. Unfortunately, conventional reconstruction methods produce poor IPD estimates, especially when conventional limited-bandwidth, limited-view handheld US probes are used for PA data collection [10, 11]. Here we focus on reconstruction methods to improve PA image quality for this geometry as a key step in delivering quantitative PA methods to the clinic.

Deep learning (DL) can significantly enhance the state-of-the-art in image reconstruction compared with conventional algorithms [12-20]. In particular, deep neural networks trained on large image datasets can learn to reconstruct images by optimizing weights in each layer from gradient descent [13]. Many recent studies have demonstrated that DL methods can greatly improve PA image quality [10, 21-32]. Generally, the accuracy of DL-reconstructed PA images can be quantified by comparing predicted with ground-truth images (e.g., calculating the peak signal-to-noise ratio (PSNR) [33, 34]). However, a major limitation to date has been the absence of validation tools in practical applications where ground-truth images are unknown. In particular, the error in DL reconstructions from conventional convolutional neural networks (CNNs) cannot be quantitatively estimated.

Researchers have developed deep-learning-based error (uncertainty) prediction methods for PA imaging. A CNN with a loss function based on the mean squared error was developed to predict different outputs for the same input using Monte Carlo dropout to estimate the uncertainty of the DL process [35]. However, it only estimated model uncertainty [17, 36], and data uncertainty [17, 36] was not considered. A different DL approach was developed to estimate errors in computed optical absorption coefficients for quantitative PA imaging [37]. An error-estimation neural network was trained with computed initial pressure images as network inputs and relative errors of the estimated optical absorption coefficients as outputs. In testing this network predicts errors in the testing dataset or real experiments. However, error estimation is not based on a more accurate statistical approach; that is, the error-estimation neural network only predicts the specific error instead of the statistical variance. In addition, to get the estimated error in computed initial pressure images (network inputs), model uncertainty should be calculated and extracted from the estimated error. Finally, the accuracy of the estimated error was not evaluated quantitatively.

A potentially more accurate way to validate PA images reconstructed from DL networks is to use a Bayesian convolutional neural network (BCNN) with a probability-distributed likelihood function as the loss function. Compared to conventional CNNs, the BCNN is an effective approach to approximate the uncertainty (pixel-wise confidence level of the image reconstruction) without knowing the ground truth. It assumes that each image pixel represents the parameter defining a probability distribution (e.g., Laplacian or Bernoulli distribution), rather than a deterministic intensity value [38]. Then, the error (uncertainty) can be quantified by Monte Carlo dropout [39] or Deep Ensembles [40]. Since the parameters of the probability distribution for each image pixel are completely predicted by the BCNN, both data and model uncertainties are estimated, and the overall uncertainty (error) is the combination of the two. It has been applied to different imaging tasks including image segmentation [41], phase imaging [17], single-pixel imaging [36], optical metrology [42], and image classification [43].

Here, we propose a hybrid BCNN (Hybrid-BCNN) providing joint PA segmentation (Bernoulli distribution) and image (Laplacian distribution) reconstructions with error prediction (i.e., overall uncertainty combining data and model uncertainties). Segmentation here denotes structural segmentation excluding artifacts and background noise. In addition to identifying PA signals, segmentation focuses the Hybrid-BCNN on minimizing the loss function only in regions of PA signals since images from a typical PAUS architecture are usually sparse, and a single probability distribution is not efficient to describe both background and PA (non-background) pixels. The proposed Hybrid-BCNN was trained on simulated PA data and makes predictions on both simulated and experimental data. At the prediction stage, besides predicting PA segmentation and image, both data and model uncertainty can be computed statistically with Monte Carlo dropout to completely estimate the error (i.e., overall uncertainty) in predicted outputs. The proposed Hybrid-BCNN was compared to another BCNN without segmentation to verify its importance and the inefficiency of a single probability distribution. As demonstrated below, the proposed Hybrid-BCNN reconstructs the PA segmentation and image with high accuracy; the segmentation helps BCNN train efficiently; and predicted PA segmentation and image errors (uncertainties) are highly statistically correlated to actual reconstruction errors from the two proposed statistical methods, indicating that it is a promising tool to make and validate DL reconstructions.

To demonstrate one potential way to leverage uncertainty predictions, confidence processing is proposed by computing the relative uncertainty to improve PA image quality (e.g., artifact removal) for both simulations and measurements. Although the proposed Hybrid-BCNN is only applied here to predictions of the IPD and its error, it can also be used in other areas of quantitative PA imaging (e.g. spectroscopic imaging).

## 2. Materials and Methods

### 2.1. BCNN theory

Rather than simple weights, BCNNs use distributions over network parameters and the training dataset [38]. That is, BCNNs assume stochastic rather than deterministic network processes (e.g., dropout [44], weight initialization [45] etc.). Denote the training dataset as $(X,Y) = \{x_n, y_n\}_{n=1}^{N}$ with $X$ and $Y$ representing network inputs and ground-truth images, respectively. N is the total number of training images. To approximate prediction variability in y given a specific input $x_{test,t}$ within $(X_{test}, Y_{test}) = \{x_{test,t}, y_{test,t}\}_{t=1}^{T}$ (T is the total number of images in the testing dataset), the predictive distribution $p(y|x_{test,t}, X, Y)$ over all possible learned weights (with marginalization) [39] is used:

$$p(y|x_{test,t}, X, Y) = \int p(y|x_{test,t}, W) p(W|X, Y) dW \tag{1}$$

where $p(y|x_{test,t}, W)$ denotes the predictive distribution including all possible output predictions given the learned weights $W$ and the input $x_{test,t}$. It represents data uncertainty [17]. $p(W|X, Y)$ denotes all possible learned weights given the training dataset, representing model uncertainty [17].

To jointly reconstruct the PA segmentation and image with uncertainty quantification, we choose the joint multivariate Bernoulli-distributed (for PA segmentation) and Laplacian-distributed (for PA image) likelihood functions to model data uncertainty, creating a Hybrid-BCNN. The specific choice of a Laplacian distribution for the image is discussed in the Supplementary Document. In this joint distribution, the Laplacian distribution is only assigned to the segmentation region, where the segmentation value is 1 under the Bernoulli distribution (detailed in Eqs. S6-S8 in the Supplementary Document).

The loss function $L_{Hybrid}(W|x,y)$ ($W$ combines network weights $W_1$ for PA segmentation and network weights $W_2$ for PA image) for the joint distributed likelihood function given the training data pair $(x_n, y_n)$, where $y_n = (y_{seg,n}, y_{image,n})$, is,

$$L_{Hybrid}(W|x_n, y_n) = \sum_{m=1}^{M}\left[(y_{seg,n}^m - 1)\log(1 - \mu_1^m) - y_{seg,n}^m \log(\mu_1^m) + y_{seg,n}^m \left(\frac{|y_{image,n}^m - \mu_2^m|}{\sigma^m} + \log(2\sigma^m)\right)\right] \quad (2)$$

where $y_{seg,n}$ and $y_{image,n}$ are the $n$th ground-truth PA segmentation and image in the training dataset, respectively. $y_{seg,n}^m$ and $y_{image,n}^m$ are the $m$th pixel of $y_{seg,n}$ and $y_{image,n}$, respectively. $\mu_1^m$ is the probability of $(y_{seg,n}^m = 1|x_n, W)$, which is also the mean of the Bernoulli-function for $y_{seg,n}^m$. $\mu_2^m$ and $\sigma^m$ are the mean and standard deviation of the Laplacian-distributed likelihood function for $y_{image,n}^m$. $M$ is the total pixel number in $y_{seg,n}$ or $y_{image,n}$. Complete derivations of Eq. (2) are presented in Supplementary Section 1.

The loss function in Eq. (2) is minimized during training. This Hybrid-BCNN has three output channels, where one ($\mu_1$) is for the Bernoulli-distributed likelihood function and two ($\mu_2$ and $\sigma$) are for the Laplacian-distributed likelihood function.

Model uncertainty is measured with a dropout network [39]. A distribution $q(W)$ (defined in ref. [39]) is learned to approximate $p(W|X,Y)$ (minimizing the Kullback-Leibler divergence between $q(W)$ and $p(W|X,Y)$) by applying a dropout layer before every layer with learnable weights. During the prediction process, model uncertainty is approximated by Monte Carlo dropout [39]. With Monte Carlo integration, the predictive distribution $p(y|x_{test,t}, X, Y)$ in Eq. (1) can be approximated as:

$$p(y|x_{test,t}, X, Y) \approx \frac{1}{K}\sum_{k=1}^{K} p(y|x_{test,t}, W^k) \quad (3)$$

where $K$ is the total number of dropout activations during prediction.

Finally, the reconstructed PA segmentation and image are represented by the predicted mean $\hat{\mu}_{test,t}^m$ (the $m$th pixel) given test input $x_{test,t}$ under Bernoulli-distributed and Laplacian-distributed likelihood functions, respectively:

$$\hat{\mu}_{test,t}^m = \frac{1}{K}\sum_{k=1}^{K} \hat{\mu}_{test,t}^{m,k} \quad (4)$$

where $\mu^m$ denotes $\mu_1^m$ or $\mu_2^m$, and $\hat{\mu}_{test,t}^{m,k}$ denotes the $m$th pixel of the predicted $\mu_1$ (for PA segmentation) or $\mu_2$ (for PA images) from the $k$th dropout activation given test input $x_{test,t}$ (i.e., $\hat{\mu}_{1test,t}^{m,k}$ or $\hat{\mu}_{2test,t}^{m,k}$).

The corresponding predicted uncertainties (the $m$th pixel) $\hat{\sigma}_{test,t(Ber)}^m$ (for PA segmentation under Bernoulli function) and $\hat{\sigma}_{test,t(Lap)}^m$ (for PA image under Laplacian function) given the test input data $x_{test,t}$ are,

$$\hat{\sigma}^m_{test,t(Ber)} = \sqrt{\frac{1}{K}\sum_{k=1}^{K}[\hat{\mu}^{m,k}_{1test,t}(1-\hat{\mu}^{m,k}_{1test,t})] + \frac{1}{K}\sum_{k=1}^{K}(\hat{\mu}^{m,k}_{1test,t} - \hat{\mu}^{m}_{1test,t})^2} \qquad (5)$$

$$\hat{\sigma}^m_{test,t(Lap)} = \sqrt{\frac{1}{K}\sum_{k=1}^{K}2(\hat{\sigma}^{m,k}_{test,t})^2 + \frac{1}{K}\sum_{k=1}^{K}(\hat{\mu}^{m,k}_{2test,t} - \hat{\mu}^{m}_{2test,t})^2} \qquad (6)$$

where $\hat{\sigma}^{m,k}_{test,t}$ denotes the predicted standard deviation of $y^m_{image}$ from the *k*th dropout activation for test data $x_{test,t}$. Complete derivations of Eq. (4-6) are presented in Supplementary Section 2.

The reconstructed PA image and its corresponding uncertainty are multiplied with the reconstructed PA segmentation since the Hybrid-BCNN only minimizes the loss function within non-background pixels (given by PA segmentation) and, therefore, the background and its uncertainty are not optimized. In this case, the accuracy (error) of PA segmentation and image predictions can be validated by the predicted uncertainties.

To show the power of the hybrid approach and to prove that a single distribution likelihood function (i.e., Laplacian) is not efficient to describe both background pixels and PA (non-background) pixels, we compared the Hybrid-BCNN to another BCNN with only the Laplacian-distributed likelihood function where there is no PA segmentation (Lap-BCNN). By taking logarithm and negative operations on Eq. (S4) in the Supplementary Document, the loss function $L_{Lap}(W_{Lap}|x,y)$ for Lap-BCNN given the training data pair $(x_n, y_n)$ where $y_n = y_{image,n}$ is,

$$L_{Lap}(W_{Lap}|x_n, y_n) = \sum_{m=1}^{M}\left[\frac{|y^m_{image,n} - \mu^m_2|}{\sigma^m} + \log(2\sigma^m)\right] \qquad (7)$$

where $W_{Lap}$ denotes the network weights of Lap-BCNN and the other variables have the same meanings as those in Eq. (2), and Eqs. (S4) and (S5) in the Supplementary Document.

The predicted PA image and uncertainty from the Lap-BCNN are the same as in Eqs. (4) and (6).

An uncertainty assessment metric [17, 46-48] is used to quantify the accuracy of uncertainty predictions in the testing dataset by computing the reliability diagram (*credibility* (**Cred**) vs *empirical accuracy* (**ACC**)). Details of computing this diagram are in Supplementary Section 3. The linear correlation coefficient (CC) between **Cred** and **ACC**, and the slope of the corresponding linear fit, are calculated to quantify the diagonality of the reliability diagram. In this paper, the bound of the credible interval $\epsilon$ equals $0.2\hat{\mu}^m_{test,t}$ to provide sufficient sample points from the discrete probability bins to appropriately plot the reliability diagram and evaluate its diagonality. To further test whether predicted uncertainties are related to true reconstruction errors, absolute reconstruction error is plotted versus 2× predicted uncertainty (i.e., 2× standard deviation). Theoretically, the absolute errors of ~95% of the plotted points should be less than or equal to their corresponding 2× predicted uncertainties.

*2.2. BCNN structure, parameters, and PA dataset simulation and preprocessing*

The Hybrid-BCNN structure implementing the computational approach presented in the last section is shown in Fig. 1, where a U-Net architecture [49] with an encoder-decoder structure with skip connections between contracting paths and expanding paths is used. The contracting path captures context and the expanding path enables precise localization [49]. Similar to the BCNN in a previous publication [36], dropout layers with a dropout rate of 0.1 appear before each convolution layer to prevent overfitting during training, and $L_2$ kernel regularizers and bias regularizers with a regularization factor of $1 \times 10^{-6}$ were included in each convolution layer. Batch normalization (Batchnorm) [50] layers appear after each convolution layer to stabilize the network and make it converge better during training, and LeakyRelu [51] is used as the activation function. For Lap-BCNN, the same architecture is used except that there are two output channels (the predicted PA image and its uncertainty) for only the Laplacian-distributed likelihood function.

In a previous publication on DL-based real-time integrated PAUS imaging [32], two preprocessing approaches were applied to data directly acquired from the transducer. One is delay-and-sum (DAS) beamforming and the other is the multiple-channel transformed array (MC), in which propagation delays were applied between all observation points in the image and each transducer element without summation across array elements [32]. The MC approach has multiple channels, where the channel number equals the transducer element number, preserving more information embedded in the original data [32]. Previous results have shown that DL predictions with MC as network inputs outperform those with DAS inputs [32]. Therefore, only MC-preprocessed data were used as network inputs in this work.

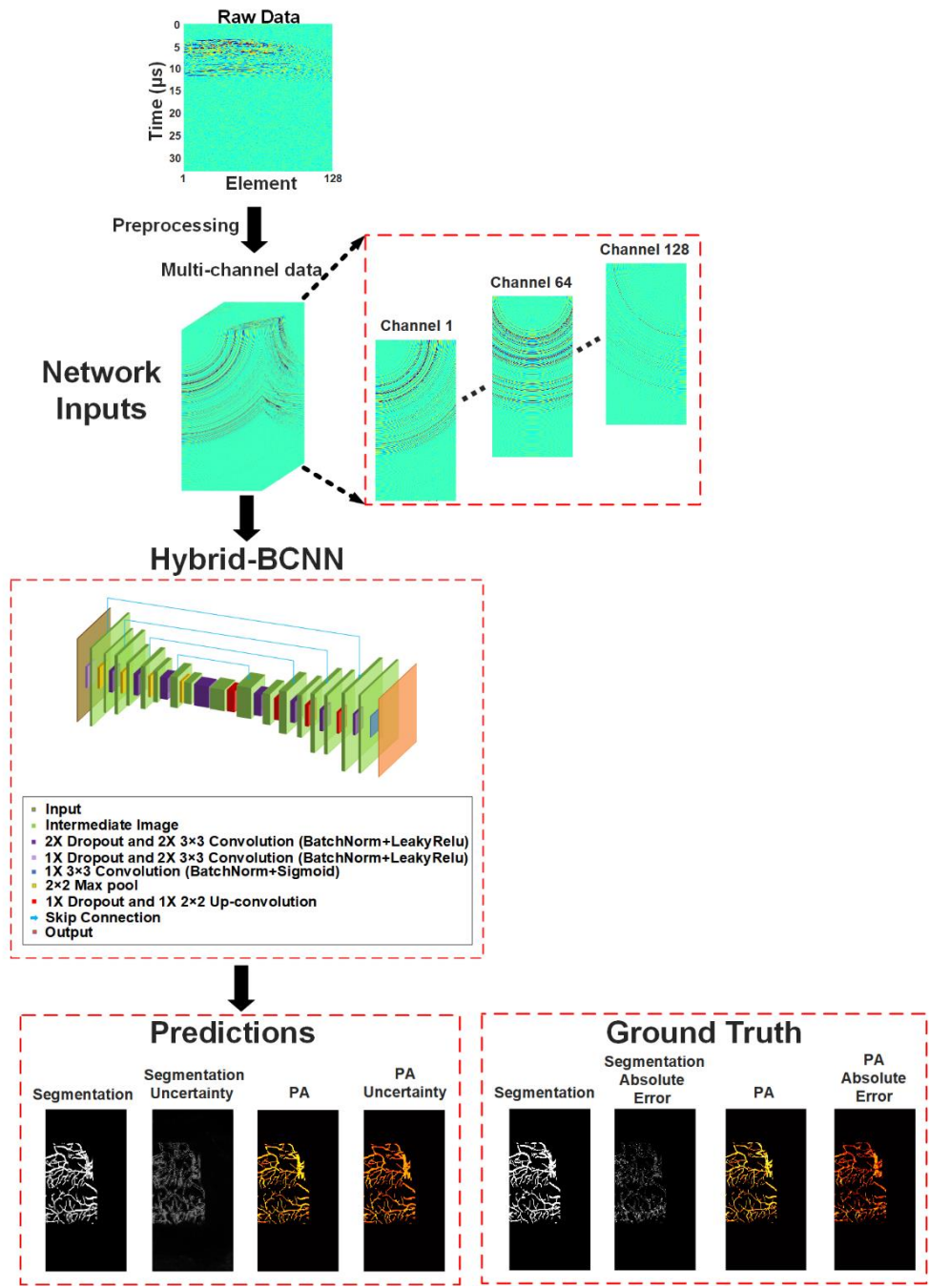

Fig. 1 The Hybrid-BCNN structure with multi-channel data (preprocessed from transducer array data) as network inputs and network predictions.

The simulated dataset was generated from the method described in a previous publication [32], with simulation parameters shown in Table S1 in Supplementary Section 4. A PA imaging forward model was created with these parameters to convert PA images to transducer array data, and a PA imaging backward model was created with these same parameters to convert transducer array data to MC data as network inputs. Note that these parameters closely mimic those for our experimental PAUS imaging system described below. PA objects in the dataset are microvessels with varying shapes, sizes and locations, which makes their signal components widely distributed in both space and frequency domains. They are representative of typical microvascular networks expected in many clinical applications. A total of 6.85% of the overall pixels in the dataset are non-background pixels (with PA signals). After shuffling the overall 16,000 images, 12,800 (80%) were used for training, 1,600 (10%) were used for validation and the remaining 1,600 (10%) were used for testing (no replicated images among the three datasets).

Table S1 presents some training hyperparameters of both BCNNs. They were trained on a NVIDIA GeForce GTX 1080 Ti GPU with 11 GB of memory. In training, the Adam optimizer was used with a constant learning rate of 0.0005 with reference to BCNN training in a previous publication [36], batch size was chosen to be 8 given limitations on the GPU memory, and the maximum epoch number was chosen to be 1,000 to guarantee complete training. An early stopping criterion was applied, where training would finish when the validation loss value did not decrease in 50 consecutive epochs (50 epochs were chosen to prevent training from stopping at incorrect places such as a plateau at the beginning of training, a sudden jump of the loss value during training, etc.), and the optimal weights with the lowest validation loss would be retrieved retrospectively. The maximum training time was approximately 35 hours.

*2.3. Experimental PAUS imaging system and phantom preparations*

To initially test the performance of the proposed BCNN for practical applications, two simple phantom objects constructed from optically absorbing metal wires with known geometries were imaged using an integrated, experimental photoacoustic/US (PAUS) system [9, 32] (as shown in Fig. 7 in [32]). The wires were twisted into the shape of the letters 'S' and 'W' and suspended in the x-z imaging plane in a cubical container. These simple shapes were used because ground-truth absorption profiles are reasonably well known. Note that data from the 'W' shaped wire was used in our previous publication of DL-based real-time integrated PAUS imaging [32]. It was reused here to test whether uncertainty predictions from the proposed Hybrid-BCNN can help with additional image interpretation. The 'S' shaped wire is new; it is used here because conventional DAS images of this object exhibit reverberation artifacts. Consequently, it can test whether the proposed Hybrid-BCNN can find, mark and, ultimately, correct the artifacts. The container was filled with a 2% intralipid solution (Fresenius Kabi, Deerfield, USA) acting as a scattering medium with effective attenuation coefficient of ~0.1 mm$^{-1}$.

The experimental PAUS system employs a unique fiber optic delivery system where each laser pulse is sequentially delivered to one of 20 different fibers. At the probe front side, the fibers are equally distributed along the azimuthal axis of the US array transducer (LA 15/128-1633, Vermon S.A. France) with 10 fibers on each side. A mechanical rotation system sequentially directs the beam from a kHz-rate, wavelength-tunable (700-900nm) diode-pumped laser (Laser Export, Russia) to a different one of the 20 fibers on each pulse, delivering laser energy to the sample at a PA frame rate of about 50 Hz. Interleaved with PA pulses were US pulses enabling simultaneous acquisition of PA and US image data at the 50 Hz frame rate (detailed in [9]). All data acquisition was controlled by a commercial US scanner (Vantage, Verasonics, WA, USA) using trigger signals created by the motor controller and encoder to ensure accurate synchronization between US and PA scan sequences.

For this study, PA data were acquired using a single wavelength at 795nm. One PA raw data frame contains 2,048 temporal samples × 128 elements × 20 fibers. Each data frame was averaged over 20 fibers to improve SNR. The size of each reconstructed PA image is 512 × 128 with an axial pixel size of 0.05 mm and lateral pixel size of 0.1 mm. Finally, PA segmentation, images and uncertainties were reconstructed from acquired experimental data using BCNNs trained with simulated data.

## 3. Results

*3.1. Simulation results and quantitative evaluation*

Hybrid-BCNN and Lap-BCNN results on two representative samples of simulated PA data are shown in Fig. 2 (a-f). Ground-truth PA segmentations, images and conventional DAS results are shown in Fig. 2 (a) and (d). The reconstructed PA segmentations and images together with their corresponding predicted uncertainties from Hybrid-BCNN are shown in Fig.2 (b) and (e), respectively. The PA images and their corresponding predicted uncertainties from Lap-BCNN are shown in Fig.2 (c) and (f), respectively. The segmentation/PA absolute error in Fig. 2 (b), (c), (e) and (f) is the absolute difference between the reconstructed segmentation/PA and the ground-truth. PA reconstructions and uncertainties are presented on a log scale over a 50 dB scale relative to the peak signal of the images in the testing dataset. PA segmentations and predicted segmentation uncertainties are presented on a linear scale.

Fig. 2(c) and (f) clearly show that the Lap-BCNN makes less accurate PA image reconstructions and uncertainty predictions. The averaged PSNR was calculated within the testing dataset (28.9855 dB for Hybrid-BCNN, 18.5948 dB for Lap-BCNN and 21.6148 dB for DAS) to quantitatively demonstrate that Lap-BCNN (as well as DAS) provides a less accurate PA image reconstruction. The reason is that PA images acquired with linear-array, limited-bandwidth ultrasound transducers are usually sparse (e.g., in this study, only 6.85% of the overall pixels in the dataset have PA signals). Therefore, the Lap-BCNN learns from pixels where the majority of their values are 0 when minimizing the loss function for the Laplacian-distributed likelihood function in Eq. (7). In this case, the Lap-BCNN focuses on the majority 0-value pixels and tends to converge to a wrong solution no matter what the network input is. Clearly, the Laplacian-distributed likelihood function alone is not sufficient to describe both background pixels and pixels with PA signals.

Therefore, we propose that a combination of Bernoulli-distributed and Laplacian-distributed likelihood functions to jointly reconstruct the PA segmentation and image (i.e., Hybrid-BCNN) is more appropriate. Segmentation using a Bernoulli-distributed likelihood function can identify background pixels and pixels with PA signals, and also focuses the network on pixels with PA signals to minimize the value of the loss function for a Laplacian-distributed likelihood function driving PA image reconstructions. Within pixel regions with PA signals there is a range of signal amplitudes, making the network learn the correct Laplacian distribution. Although pixel values (background pixels vs PA pixels) are still unevenly distributed for segmentation, the Laplacian term $y_{seg,n}^m (\frac{|y_{image,n}^m - \mu_2^m|}{\sigma^m} + \log(2\sigma^m))$ will balance the Bernoulli term $(y_{seg,n}^m - 1)\log(1 - \mu_1^m) - y_{seg,n}^m \log(\mu_1^m)$ in Eq. (2) to learn the correct segmentation.

As shown in Fig. 2 (b) and (e), both the PA segmentation and image, along with corresponding uncertainties, are reconstructed with Hybrid-BCNN. The final segmentation was computed by thresholding the predicted segmentation at the mid-range value of 0.5 (i.e., pixel values larger than 0.5 are rounded up to 1 and remaining pixel values are rounded down to 0). The low segmentation/PA absolute errors show that these predictions are very accurate. Uncertainty predictions by the Hybrid-BCNN of PA segmentation and image are highly statistically correlated to the segmentation/PA absolute errors, as quantitatively shown below.

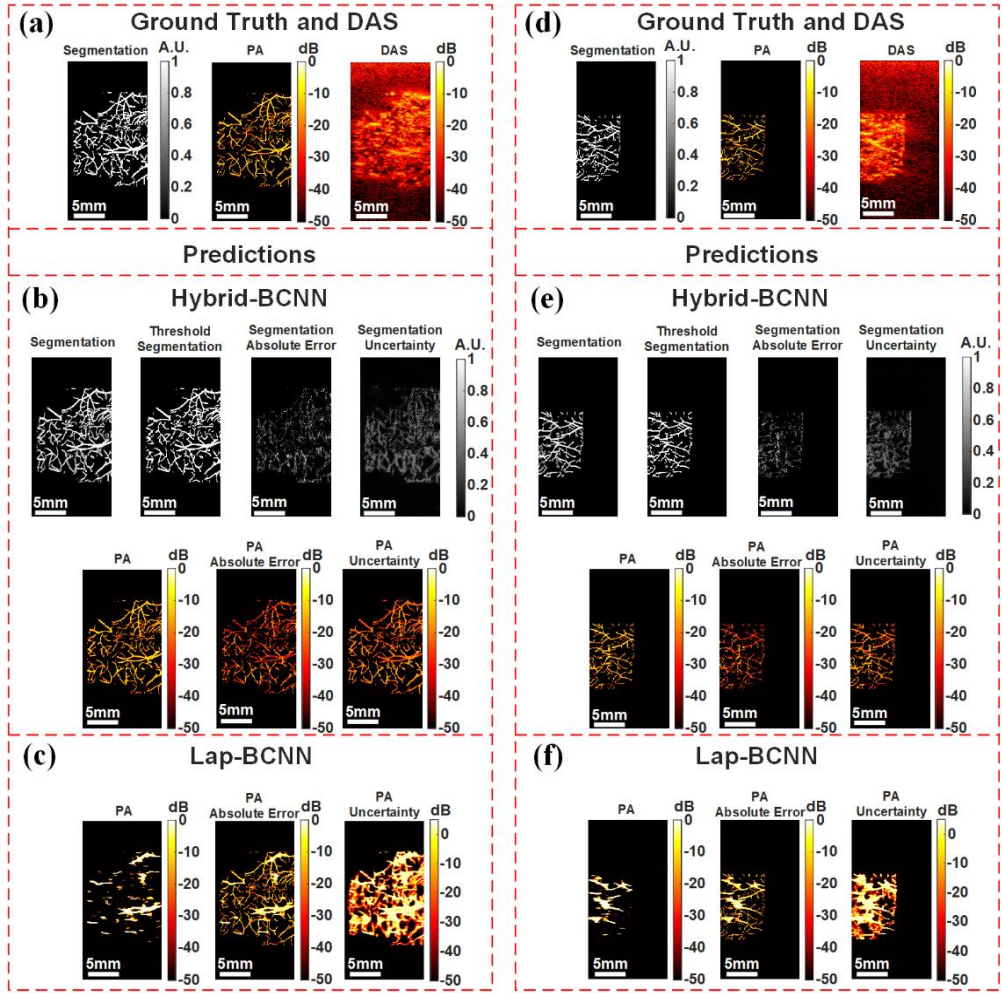

Fig. 2 Hybrid-BCNN and Lap-BCNN results for simulated PA data. (a) A representative ground-truth PA segmentation and image, and the corresponding DAS-reconstructed PA image. (b) Reconstructions and prediction errors from Hybrid-BCNN. (c) Reconstructions and prediction errors from Lap-BCNN. (d-f) are the same as (a-c) except for a different image sample. The scale bars denote 5 mm.

The credibility map (Cred Map) and reliability diagram (ACC vs Cred) were computed for the results in Fig. 2 and presented in Fig. 3. They are highly statistically correlated to the PA absolute errors in Fig. 2 (b) and (e), where pixels with low PA absolute errors are marked with high credibility. By computing the CC and slopes of the linear fits, the reliability diagrams from the Hybrid-BCNN in Fig. 3 show high levels of diagonality. These results indicate that the predicted uncertainties from the Hybrid-BCNN match well with the absolute errors and can be a reliable tool to quantify reconstruction errors.

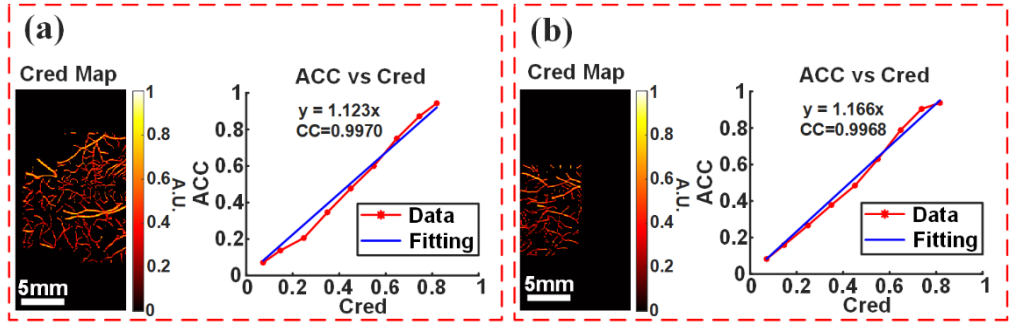

Fig. 3 Evaluation of Hybrid-BCNN PA image uncertainty predictions with the Cred Map (credibility) and ACC vs Cred (reliability diagram) calculation. (a) The Cred Map and ACC vs Cred (with CC and the slope of the linear fitting line) calculation in Hybrid-BCNN for the image sample in Fig. 2. (a-c). (b) is the same as (a) except that the results are for the image sample in Fig. 2(d-f). The scale bars denote 5 mm.

To further demonstrate the strong relationship between predicted uncertainties and reconstruction errors, absolute errors vs 2× predicted uncertainties (i.e., 2× standard deviation) results were computed for Fig. 2 and presented in Fig. 4. Figure 4 (a) and (b) correspond to the results in Fig. 2(b), and Fig. 4 (c) and (d) correspond to the results in Fig. 2(e). Fig. 4(a) shows that 98.14% of the overall pixels (within the predicted segmentation regions) have absolute errors less than or equal to 2× predicted uncertainties (i.e., 2× standard deviation) for this case. As shown in Fig. 4(b), within a specific range of 2× predicted uncertainties [95% × ½ max(2×uncertainty), 105% × ½ max(2×uncertainty)], 96.58% of the pixels have absolute errors less than or equal to 2× predicted uncertainties. This indicates that the predicted PA image uncertainty in Fig. 2(b) can accurately quantify the errors in the reconstructed PA image. The same conclusion can be made from Fig. 4(c) and (d) for the results in Fig. 2(e).

Quantitative metrics (segmentation accuracy, PSNR, segmentation CC, CC and slopes of the ACC vs Cred reliability diagrams) were computed for all samples in the testing dataset to evaluate the Hybrid-BCNN, as shown in Table S2 in Supplementary Section 5 (average value and standard deviation (in the parentheses) for each metric). The results show that, quantitatively, the Hybrid-BCNN can accurately reconstruct PA segmentations, images and corresponding uncertainties. In addition, the Hybrid-BCNN reconstructs comparable PA images as conventional CNNs (as shown in Table III in [32]) with a similar PSNR.

The selection of probability distribution functions was also studied and the results are presented in Supplementary Section 5. For segmentation, the Bernoulli-distributed likelihood function is the only practical option. However, for PA image reconstruction within vessel regions, there are multiple options. Here, we compare Hybrid-BCNNs with Laplacian-distributed and Gaussian-distributed likelihood functions. The loss function for Hybrid-BCNN with the Gaussian-distributed likelihood function is derived in detail in Supplementary Section 5. The results show that the Hybrid-BCNN is robust to the specific probability distribution function for PA image reconstruction and its uncertainty as long as the selected function is reasonable. The choice of the Laplacian-distributed likelihood function here is because of its slightly better performance in simulations.

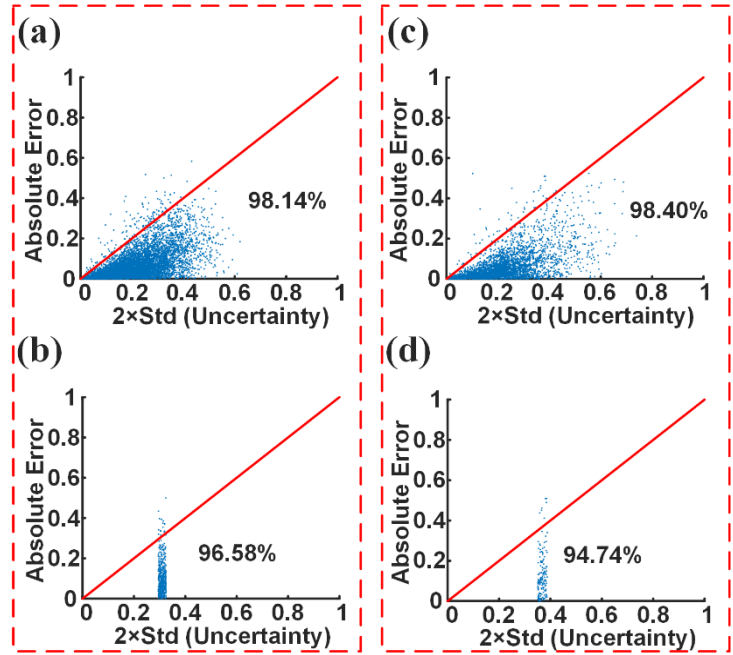

Fig. 4 Absolute Error vs 2× predicted uncertainty (i.e., 2× standard deviation (std)). (a) Absolute Error vs 2× predicted uncertainty for all pixels (within reconstructed segmentation regions) corresponding to the results in Fig. 2(b). (b) Absolute Error vs 2× predicted uncertainty for pixels within a specific range of 2× predicted uncertainties [95% × ½ max(2×uncertainty), 105% × ½ max(2×uncertainty)] from (a). (c) Absolute Error vs 2× predicted uncertainty for all pixels (within reconstructed segmentation regions) corresponding to the results in Fig. 2(e). (d) Absolute Error vs 2× predicted uncertainty for pixels within a specific range of 2× predicted uncertainties [95% × ½ max(2×uncertainty), 105% × ½ max(2×uncertainty)] from (c). Red lines indicate absolute error = 2×predicted uncertainty (std).

## 3.2. Confidence processing

The strong statistical correlation between absolute errors and predicted uncertainties suggests that the predicted uncertainty can help determine the confidence in the reconstructed PA image. For example, consider a pixel with a high signal (segmentation or PA value) and low uncertainty such that the relative uncertainty (standard deviation normalized to the mean) is small. There is high confidence in the reconstructed value for this pixel. In contrast, for a pixel with a low mean and a high uncertainty (high relative uncertainty), there is low confidence in the reconstructed value. This information can help identify artifacts and improve image quality. In general, confidence may provide a tool to judge the accuracy of quantitative PA images.

Here we explore a straightforward processing method to enhance PA images using confidence in the reconstruction, where the confidence is simply related to the ratio of the standard deviation to the mean (SD/M) at each pixel output by the Hybrid-BCNN. High confidence corresponds to a low value of this ratio. As a simple example of how to use this information, we set a threshold on the SD/M to eliminate pixels with low confidence. This operation acts like the segmentation threshold but uses relative uncertainties of both the segmentation and PA image, not just the mean value of the segmentation. The threshold can be varied to present images at different confidence levels.

Figure 5 shows this approach in detail using the image and results from Fig. 2 (a) and (b). The segmentation and segmentation uncertainty results are copied from Fig. 2 (b) and shown in Fig. 5 (a) and (b). The relative segmentation uncertainty in Fig. 5 (c) is generated by calculating the pixel-wise SD/M ratio using the results in Fig. 5 (b) and (a). Note that the predicted segmentation is not exactly zero at all pixels with no PA signal due to the stochastic nature of the network, and in those pixels the SD/M ratio is quite variable. To eliminate all pixels from further consideration that clearly do not have a PA signal, a soft threshold (0.05 used here – results are not significantly different for any value less than 0.1) is first applied. Pixels with a predicted segmentation larger than the threshold are retained to produce the final relative segmentation uncertainty image (Fig. 5 (c)).

The final confident segmentation presented in Fig. 5 (d) was then generated by thresholding the segmentation based on the SD/M ratio (all pixels with an SD/M $< 1$ are retained for the example shown here) and setting the final value to one if the segmentation value is $> 0.5$ and to zero if the value is $\leq 0.5$. This approach follows the procedure described in the last section but adds a level of confidence to the final segmentation.

Similarly, the relative PA uncertainty in Fig. 5 (g) is the pixel-wise SD/M ratio using Fig. 5 (f) and (e) within the non-zero pixel regions in Fig. 5 (d). Finally, the confident PA image in Fig. 5 (h) is generated by eliminating (i.e., set to 0) pixels in Fig. 5 (e) with corresponding SD/M above the threshold (a value of 0.9 is used here). The ground-truth PA image in Fig. 5 (i) (copied from Fig. 2 (a)) is compared with the confident PA image. This processing can eliminate obvious image artifacts but the final reconstructed image is a function of the selected threshold. Clearly, a user can vary the threshold in real-time to see images at different confidence levels to aid image interpretation. This approach is explored further using the experiments presented below.

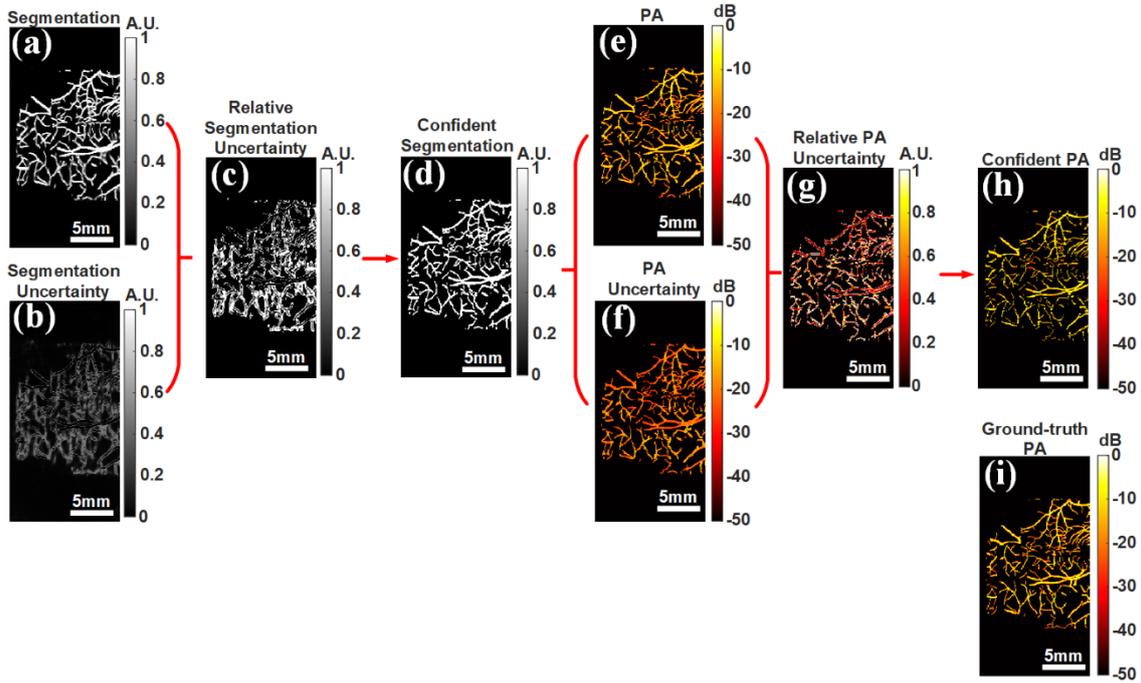

Fig. 5 Illustration of confidence processing. (a) Predicted segmentation. (b) Predicted segmentation uncertainty. (c) Relative segmentation uncertainty (SD/M) generated from (a, b). (d) Confident segmentation generated from (a, c) with a specific confidence level (i.e., relative segmentation uncertainty < 1). (e) Predicted PA image (multiplied with (d)). (f) Predicted PA uncertainty (multiplied with (d)). (g) Relative PA uncertainty (SD/M) generated from (e, f). (h) Confident PA image generated from (e, g) with a specific confidence level (i.e., relative PA uncertainty ≤ 0.9). (i) Ground-Truth PA image for reference. The scale bars denote 5mm.

## 3.3. Experimental results

Experimental results of the two wire phantom objects described in *Section 2.3* are presented here. Figure 6 shows the conventional DAS reconstruction results and Hybrid-BCNN results for the 'W' and 'S' shaped objects using exactly the same procedure as that used to produce the simulation results in Fig. 2 except that a soft segmentation threshold of 0.05 is first used to generate the corresponding PA and PA uncertainty images. The Lap-BCNN results are not shown since, as demonstrated above, they provide less accurate PA images. PA reconstructions and uncertainties are presented on a 50 dB log scale relative to the peak signal in the image. PA segmentations and predicted segmentation uncertainties are presented on a linear scale.

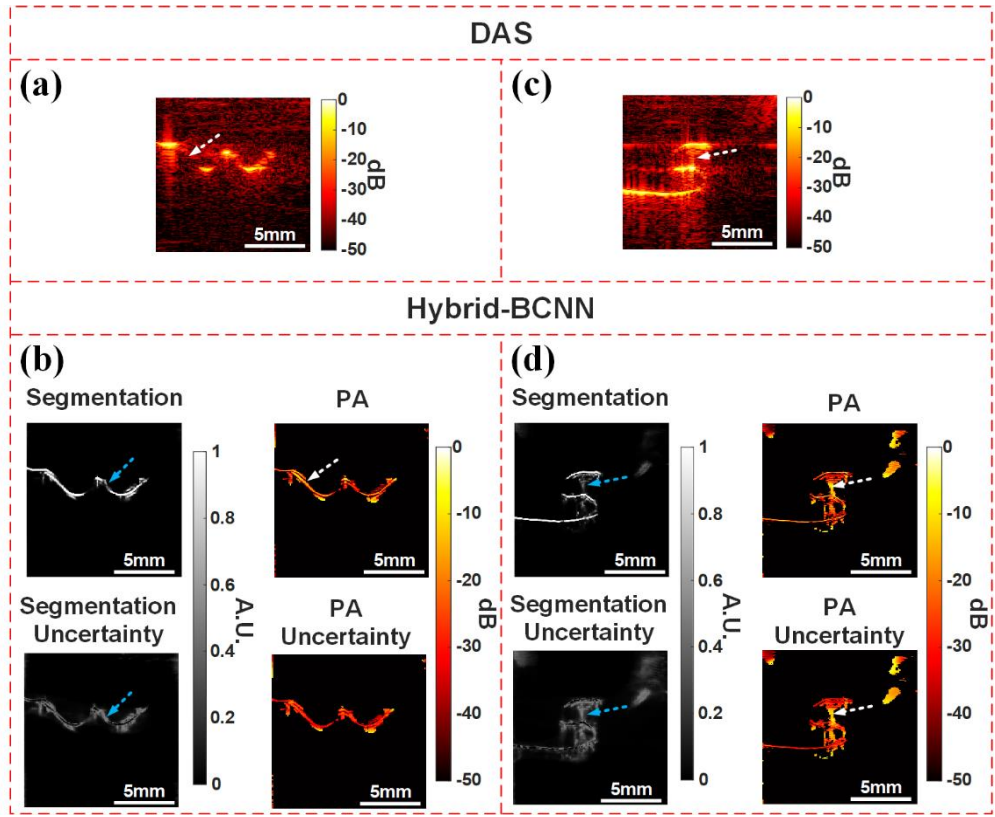

Fig. 6 Hybrid-BCNN results for experimental PA data ('S' and 'W' shaped objects). (a) DAS reconstruction of the 'W' shape object. (b) Results from Hybrid-BCNN for the 'W' shaped object. (c, d) are the same as (a, b) except for the 'S' shaped object. The scale bars denote 5 mm.

Similar to the findings in the simulation, Hybrid-BCNN outperforms DAS in PA image reconstruction, as demonstrated in Fig. 6 (a) and (b) where the local structure of the W-shaped wire indicated by the white dashed arrows can be reconstructed by Hybrid-BCNN but is absent in the DAS image. Note that the Hybrid-BCNN provides a comparable PA image to conventional CNNs [32] for the W-shaped wire, but the added uncertainty information highlights most missing features.

Although the ground-truth is only approximately known for these objects, most obvious artifacts and missing features can be easily identified and compared to uncertainty predictions to test whether they can be used to identify these errors even when the ground truth is unknown (e.g., in clinical imaging). As presented in Fig. 6 (b), the missing feature marked by the blue dashed arrow in the predicted PA segmentation is also marked with high uncertainties by the corresponding blue dashed arrow in the predicted PA segmentation uncertainty. The reverberation artifacts marked by the blue dashed arrow in the predicted PA segmentation shown in Fig. 6 (d) are also marked with high uncertainties by the corresponding blue dashed arrow in the predicted PA segmentation uncertainty. For the PA image, the reverberation artifacts marked by the white dashed arrow in the image shown in Fig. 6 (d) are still marked with high uncertainties by the white dashed arrow in the predicted PA image uncertainty.

The results in Fig. 6 were also used to test confidence processing when the ground truth is not precisely defined using the same approach as in Fig. 5. The relative segmentation uncertainty threshold is 1, as in the simulations. The confident PA images of the 'W' and 'S' shaped objects at different PA confidence levels (i.e., SD/M thresholds) are presented in Fig. 7 (a-h). As the confidence level increases (i.e., decreasing SD/M), more artifacts are removed, as expected. For the 'S' shaped object, in particular, reverberation artifacts from the original predicted PA image as shown by white arrows in Fig. 6 (d) are removed in Fig. 7 (e-h) at all 4 confidence levels. However, at very low levels of the SD/M threshold, useful image information is also removed. Therefore, there is a tradeoff between artifact removal and loss of real image features using this simple confidence processing approach. In general, a user can tune the threshold to see images at different confidence levels for different imaging tasks.

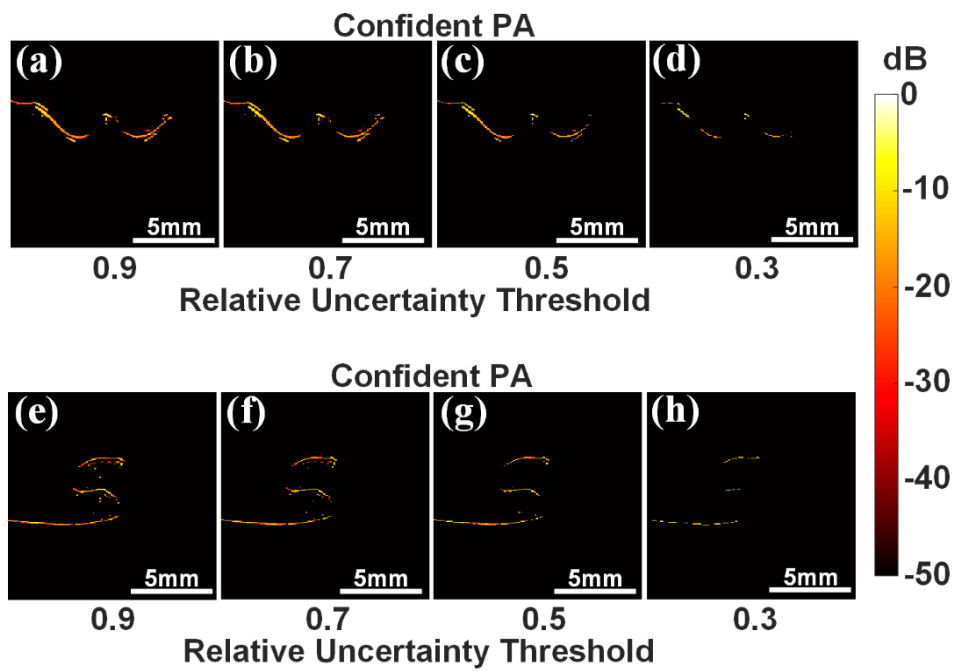

Fig. 7 Confident PA images with different confident levels (i.e., relative uncertainty thresholds) for the 'W' and 'S' shaped objects. (a-d) Results for the 'W' shaped objects. (e-h) Results for the 'S' shaped object. The scale bars denote 5mm.

Overall, these experimental results suggest that the proposed Hybrid-BCNN has the potential to be a reliable tool to make comparable DL predictions as conventional CNNs and, at the same time, at least partially validate DL reconstructions from predicted uncertainties when the ground-truth is unknown (e.g., DL-reconstructed clinical images). The proposed approach can help to evaluate reconstruction confidence, identify artifacts and, potentially, remove them with simple processing. This can be a very useful tool to validate quantitative PA methods and provide a measure of the confidence in quantitative results. Clearly, more sophisticated approaches can be used, but the relative uncertainties output by this network can help separate artifacts from image features.

## 4. Discussion

We presented predicted uncertainties to quantify errors in PA segmentations and images and used these predictions to generate confident PA images. However, we have not fully used these uncertainties to improve network predictions themselves. Therefore, our future work will focus on using predicted uncertainties as feedback to tune network structures or parameters and better select training datasets. We will also focus on applying the proposed BCNN to *in vivo* PA studies to explore how uncertainty predictions can help clinical applications of DL-based quantitative PA imaging.

Although the proposed BCNN is used to predict the IPD and its error in this paper, it can also improve quantitative PA imaging since the error prediction of the IPD can help produce a more reliable reconstruction of the optical absorption. As a further step, we can extend the BCNN approach to directly predict the optical absorption coefficient and its error, and, therefore, the confidence in the predicted absorption coefficient.

The simple confidence processing approach presented here leverages the strong correlation between predicted uncertainties and actual errors to improve PA image quality (e.g., artifact removal). Although it is very effective at removing artifacts, it is far from perfect because it can also eliminate real image features. A human observer can vary the confidence level of the PA image to tune the tradeoff between artifact and true feature removal for a given imaging task, but more sophisticated schemes should be explored to optimize this tradeoff. We also will investigate other ways to use confidence measures for different applications. For example, confidence processing can help select high confidence pixels in quantitative PA images of blood vessels to determine blood oxygenation levels. Given the typical artifacts in limited view and bandwidth imaging, high confidence pixels are much more likely to produce more accurate blood oxygenation values for real-time PAUS imaging with handheld US arrays.

Finally, the MC approach was used to preprocess measurement data to facilitate network training. We would also like to explore unprocessed measurement data as BCNN inputs. We expect that a much deeper neural network and a much larger training dataset are needed since this BCNN also must learn to map from measurement to image domains. However, the added information in unprocessed signals may lead to more robust images and uncertainty predictions.

## 5. Conclusions

In summary, we proposed a Hybrid-BCNN for limited-view-and-bandwidth PA imaging to jointly reconstruct PA segmentations (structural segmentation excluding artifacts and background noise) and images with uncertainty quantification by minimizing the loss function in Eq. (2), which jointly combines Bernoulli-distributed and Laplacian-distributed likelihood functions. Compared to conventional CNNs, BCNNs not only reconstruct images but also predict uncertainties in these images without knowing the ground truth. Due to the sparsity of PA images acquired in limited-view-and-bandwidth PA imaging (e.g., with conventional handheld linear-array transducers), a single probability distribution is not sufficient to describe both background and PA (non-background) pixels. Therefore, two probability distributions (Bernoulli and Laplacian) are used, and the segmentation focuses the network on minimizing the loss function in regions with PA signals. In addition, it can identify background regions and PA signal regions.

Simulation and experimental results show that the Hybrid-BCNN accurately reconstructs PA segmentations, images and corresponding uncertainties. Compared to the poor results from Lap-BCNN, which only reconstructs PA images without segmentation, the importance of segmentation in Hybrid-BCNN is verified in simulations. Predicted uncertainties were also shown to strongly statistically correlate with actual reconstruction errors with two statistical methods. Thus, uncertainty predictions provide a potentially reliable tool to validate PA segmentations and images by marking incorrect reconstruction areas (e.g., missing features and artifacts) with high uncertainty. This can be very important in practical applications (e.g., clinical imaging) where the ground truth is unknown. Confidence processing is effective at removing artifacts to further improve PA image quality. With some adjustments, the proposed Hybrid-BCNN and confidence processing can be applied directly to quantitative PA imaging. In particular, confidence processing may be a powerful tool to accurately estimate blood oxygenation from spectroscopic PA data acquired with conventional handheld ultrasound array transducers.

## 6. Appendix A. Supplementary Materials

Supplementary material related to this article can be found online at: link will be provided.

## 7. CRediT Authorship Contribution Statement

**Ruibo Shang**: Conceptualization, Data curation, Formal analysis, Investigation, Methodology, Software, Validation, Visualization, Writing – original draft, Writing – review & editing.

**Geoffrey P. Luke**: Conceptualization, Methodology, Writing – review & editing.

**Matthew O'Donnell**: Conceptualization, Funding acquisition, Methodology, Project administration, Resources, Supervision, Writing – review & editing.

## 8. Declaration of Competing Interests

The authors declare that they have no known competing financial interests or personal relationships that could have appeared to influence the work reported in this paper.

## 9. Data and Code Availability

Data and code underlying the results presented in this paper are not publicly available at this time but may be obtained from the authors upon reasonable request.

## 10. Acknowledgements

This work was supported by the National Institutes of Health (R01EB030484). The authors acknowledge Professor MinWoo Kim from Pusan National University for providing simulated and experimental PA dataset, and Professor Ivan M. Pelivanov from University of Washington for valuable discussions related to the results.

# Supplementary Document: Joint Segmentation and Image Reconstruction with Error Prediction in Photoacoustic Imaging using Deep Learning

Ruibo Shang[1], Geoffrey P. Luke[2], Matthew O'Donnell[1,*]

[1]uWAMIT Center, Department of Bioengineering, University of Washington, Seattle, WA 98195, USA
[2]Thayer School of Engineering, Dartmouth College, Hanover, NH 03755, USA

*odonnel@uw.edu

This document provides supplementary information to "Joint Segmentation and Image Reconstruction with Error Prediction in Photoacoustic Imaging using Deep Learning". We provide more details including: complete derivations of equations in the main text, parameters for the simulated photoacoustic (PA) data and the Bayesian convolutional neural networks (Hybrid-BCNN and Lap-BCNN), and comparison between Hybrid-BCNNs with Laplacian-distributed (used in the main text) and Gaussian-distributed likelihood functions.

## 1. Derivations of the loss function in Hybrid-BCNN

We define the Bernoulli-distributed likelihood function as,

$$p_{Ber}(y_{seg}|x, W_1) = \prod_{m=1}^{M} p_{Ber}(y_{seg}^m|x, W_1) \tag{S1}$$

$$p_{Ber}(y_{seg}^m = 1|x, W_1) = \mu_1^m \tag{S2}$$

$$p_{Ber}(y_{seg}^m|x, W_1) = (\mu_1^m)^{y_{seg}^m}(1 - \mu_1^m)^{1-y_{seg}^m} \tag{S3}$$

where $y_{seg}$ denotes the predicted PA segmentation, $y_{seg}^m$ is the $m$th pixel in $y_{seg}$, $W_1$ are network weights, $x$ is the network input, $M$ is the total pixel number in $y_{seg}$, and $\mu_1^m$ is the probability of $(y_{seg}^m = 1|x, W)$, which is also the mean of the Bernoulli-function for $y_{seg}^m$.

The multivariate Laplacian-distributed likelihood function is:

$$p_{Lap}(y_{image}|x, W_2) = \prod_{m=1}^{M} p_{Lap}(y_{image}^m|x, W_2) \tag{S4}$$

$$p_{Lap}(y_{image}^m|x, W_2) = \frac{1}{2\sigma^m}\exp\left(-\frac{|y_{image}^m - \mu_2^m|}{\sigma^m}\right) \tag{S5}$$

where $y_{image}$ is the predicted PA image, $y_{image}^m$ is the $m$th pixel in $y_{image}$, $W_2$ are network weights, $x$ is the network input, $M$ is the total pixel number in $y_{image}$, and $\mu_2^m$ and $\sigma^m$ are the mean and standard deviation of the Laplacian-distributed likelihood function for $y_{image}^m$. Note that $M$ is the same as that in Eq. (S1) since $y_{image}$ and $y_{seg}$ have the same size.

According to Eqs. (S1-S5), the joint multivariate Bernoulli-distributed (for PA segmentation) and Laplacian-distributed (for PA image) likelihood function can be defined as,

$$p_{joint} = p_{Ber}(y_{seg}|x, W_1)p_{Lap}(y_{image}|x, W_2, y_{seg}) = \prod_{m=1}^{M} p_{Ber}(y_{seg}^m|x, W_1)p_{Lap}(y_{image}^m|x, W_2, y_{seg}^m) \quad (S6)$$

$$p_{Lap}(y_{image}^m|x, W_2, y_{seg}^m = 1) = \frac{1}{2\sigma^m} \exp\left(-\frac{|y_{image}^m - \mu_2^m|}{\sigma^m}\right) \quad (S7)$$

$$p_{Lap}(y_{image}^m|x, W_2, y_{seg}^m = 0) = \emptyset \quad (S8)$$

where all variables are the same as those in Eqs. (S1-S5). Specifically, $W_1$ and $W_2$ are partial network weights for the PA segmentation and image, respectively. Eqs. (S6-S8) ensure that only non-background (i.e., finite PA signal) pixels ($y_{seg}^m = 1$) are used for PA image reconstructions and uncertainty quantification.

By taking logarithm and negative operations on Eq. (S6), the loss function $L_{Hybrid}(W|x, y)$ ($W$ combines $W_1$ and $W_2$) for the joint distributed likelihood function given the training data pair ($x_n, y_n$), where $y_n = (y_{seg,n}, y_{image,n})$, is,

$$L_{Hybrid}(W|x_n, y_n) = \sum_{m=1}^{M}\left[(y_{seg,n}^m - 1)\log(1 - \mu_1^m) - y_{seg,n}^m \log(\mu_1^m) + y_{seg,n}^m\left(\frac{|y_{image,n}^m - \mu_2^m|}{\sigma^m} + \log(2\sigma^m)\right)\right] \quad (S9)$$

where $y_{seg,n}$ and $y_{image,n}$ are the *n*th ground-truth PA segmentation and image in the training dataset, respectively. $y_{seg,n}^m$ and $y_{image,n}^m$ are the *m*th pixel of $y_{seg,n}$ and $y_{image,n}$, respectively. The other variables are the same as in Eqs. (S6-S8). Eq. S9 is Eq. (2) in the main text.

## 2. Derivations of the PA segmentation, image and corresponding uncertainties in Hybrid-BCNN

The reconstructed PA segmentation and image are represented by the predicted mean $\hat{\mu}_{test,t}^m$ (the *m*th pixel) given test input $x_{test,t}$ under Bernoulli-distributed and Laplacian-distributed likelihood functions, respectively:

$$\hat{\mu}_{test,t}^m = \mathbb{E}[\mu^m|x_{test,t}, X, Y] \approx \frac{1}{K}\sum_{k=1}^{K}\mathbb{E}[\mu^m|x_{test,t}, W^k] \approx \frac{1}{K}\sum_{k=1}^{K}\hat{\mu}_{test,t}^{m,k} \quad (S10)$$

where $\mathbb{E}$ denotes the expectation, $\mu^m$ denotes $\mu_1^m$ or $\mu_2^m$, and $\hat{\mu}_{test,t}^{m,k}$ denotes the *m*th pixel of the predicted $\mu_1$ (for PA segmentation) or $\mu_2$ (for PA images) from the *k*th dropout activation given test input $x_{test,t}$ (i.e., $\hat{\mu}_{1test,t}^{m,k}$ or $\hat{\mu}_{2test,t}^{m,k}$).

For segmentation, the predicted uncertainty $\hat{\sigma}_{test,t(Ber)}^m$ of the *m*th pixel given the test input data $x_{test,t}$ under the Bernoulli-distributed likelihood function is,

$$\hat{\sigma}_{test,t(Ber)}^m = \sqrt{Var(y_{seg}^m|x_{test,t}, X, Y)} = \sqrt{\mathbb{E}[Var(y_{seg}^m|x_{test,t}, W)] + Var(\mathbb{E}[y_{seg}^m|x_{test,t}, W])}$$

$$\approx \sqrt{\frac{1}{K}\sum_{k=1}^{K}[\hat{\mu}_{1test,t}^{m,k}(1-\hat{\mu}_{1test,t}^{m,k})] + \frac{1}{K}\sum_{k=1}^{K}(\hat{\mu}_{1test,t}^{m,k} - \hat{\mu}_{1test,t}^{m})^2} \tag{S11}$$

where $Var$ denotes pixel-wise variance, $\hat{\sigma}_{test,t(Ber)}^{m(D)} = \sqrt{\frac{1}{K}\sum_{k=1}^{K}[\hat{\mu}_{1test,t}^{m,k}(1-\hat{\mu}_{1test,t}^{m,k})]}$ denotes data uncertainty and $\hat{\sigma}_{test,t(Ber)}^{m(M)} = \sqrt{\frac{1}{K}\sum_{k=1}^{K}(\hat{\mu}_{1test,t}^{m,k} - \hat{\mu}_{1test,t}^{m})^2}$ denotes model uncertainty.

The predicted uncertainty $\hat{\sigma}_{test,t(Lap)}^{m}$ of the $m$th pixel for test data $x_{test,t}$ for the Laplacian-distributed likelihood function is,

$$\hat{\sigma}_{test,t(Lap)}^{m} = \sqrt{Var(y_{image}^{m}|x_{test,t}, X, Y)} = \sqrt{\mathbb{E}[Var(y_{image}^{m}|x_{test,t}, W)] + Var(\mathbb{E}[y_{image}^{m}|x_{test,t}, W])}$$

$$\approx \sqrt{\frac{1}{K}\sum_{k=1}^{K}2(\hat{\sigma}_{test,t}^{m,k})^2 + \frac{1}{K}\sum_{k=1}^{K}(\hat{\mu}_{2test,t}^{m,k} - \hat{\mu}_{2test,t}^{m})^2} \tag{S12}$$

where $Var$ denotes pixel-wise variance, $\hat{\sigma}_{test,t}^{m,k}$ denotes the predicted standard deviation of $y_{image}^{m}$ from $k$th dropout activation for test data $x_{test,t}$. $\hat{\sigma}_{test,t(Lap)}^{m(D)} = \sqrt{\frac{1}{K}\sum_{k=1}^{K}2(\hat{\sigma}_{test,t}^{m,k})^2}$ denotes data uncertainty, and $\hat{\sigma}_{test,t(Lap)}^{m(M)} = \sqrt{\frac{1}{K}\sum_{k=1}^{K}(\hat{\mu}_{2test,t}^{m,k} - \hat{\mu}_{2test,t}^{m})^2}$ denotes model uncertainty.

## 3. Reliability diagram

An uncertainty assessment metric [1-4] is used to quantify the accuracy of uncertainty predictions in the testing dataset. The probability density $f_{test,t}^{m}(a)$ (Laplacian distribution model) of $\hat{y}_{test,t}^{m}$ (the $m$th pixel of the image reconstructed from $x_{test,t}$) for the value $a$ is,

$$f_{test,t}^{m}(a) = p(\hat{y}_{test,t}^{m} = a|x_{test,t}, X, Y) \approx \frac{1}{K}\sum_{k=1}^{K}p_{Lap}(a; \hat{\mu}_{test,t}^{m,k}, \hat{\sigma}_{test,t}^{m,k}) \tag{S13}$$

where $p_{Lap}$, $\hat{\mu}_{test,t}^{m,k}$, $\hat{\sigma}_{test,t}^{m,k}$ and $K$ are the same as in Eqs. (S5), (S10) and (S12).

Then, a credible interval is defined as $[\hat{\mu}_{test,t}^{m} - \epsilon, \hat{\mu}_{test,t}^{m} + \epsilon]$ where $\hat{\mu}_{test,t}^{m}$ is given in Eq. (S10) and $\epsilon$ is the bound. Therefore, the *credibility* $c_{test,t}^{m,\epsilon}$, the probability that $a$ falls within the credible interval, is,

$$c_{test,t}^{m,\epsilon} = \int_{\hat{\mu}_{test,t}^{m} - \epsilon}^{\hat{\mu}_{test,t}^{m} + \epsilon} f_{test,t}^{m}(a)da = \frac{1}{K}\sum_{k=1}^{K}[F^k(\hat{\mu}_{test,t}^{m} + \epsilon) - F^k(\hat{\mu}_{test,t}^{m} - \epsilon)] \tag{S14}$$

where $F^k$ is the cumulative distribution function of the predicted Laplacian distribution from the $k$th Monte Carlo Dropout.

To quantitatively evaluate the *credibility*, $c_{test,t}^{m,\epsilon}$ is compared with the *empirical accuracy*, which is the empirical probability that the ground truth PA value matches the reconstructed PA value by computing the reliability diagram [1]. If the predicted uncertainty is accurate, the *credibility* should be similar to the *empirical accuracy* (i.e., the reliability diagram is diagonal). To divide the reliability diagram into discrete probability bins, the average *credibility* and *empirical accuracy* are calculated. With H probability bins (i.e., the bin interval is 1/H), the *h*th bin $B_h$ is bounded by $(h-1)/H$ and $h/H$. The average *credibility* is the *credibility* averaged over the set of pixels $S_h^\epsilon$ whose *credibility* drops within $((h-1)/H, h/H]$,

$$\boldsymbol{Cred}(B_h, \epsilon) = \frac{1}{|S_h^\epsilon|} \sum_{c_{test,t}^{m,\epsilon} \in ((h-1)/H, h/H]}^{m \in [1,M]} c_{test,t}^{m,\epsilon} \tag{S15}$$

where $|S_h^\epsilon|$ denotes the total number of pixels within $((h-1)/H, h/H]$. The other variables are the same as in Eqs. (S4) and (S14).

The *empirical accuracy (ACC)* is defined as the fraction of the pixels in set $S_h^\epsilon$ where the ground truth PA image pixel value $y_{test,t}^m$ ($m \in S_h^\epsilon$) is indeed within the credible interval $[\hat{\mu}_{test,t}^m - \epsilon, \hat{\mu}_{test,t}^m + \epsilon]$,

$$\boldsymbol{ACC}(B_h, \epsilon) = \frac{1}{|S_h^\epsilon|} \sum_{\hat{\mu}_{test,t}^m - \epsilon \leq y_{test,t}^m \leq \hat{\mu}_{test,t}^m + \epsilon}^{m \in S_h^\epsilon} (+1) \tag{S16}$$

where the variables are the same as those in Eqs. (S10) and (S15). Note that $(+1)$ is to count pixels.

Finally, the linear correlation coefficient (CC) between $\boldsymbol{Cred}(B_h, \epsilon)$ and $ACC(B_h, \epsilon)$, and the slope of the corresponding linear fit, are calculated to quantify the diagonality of the reliability diagram. In this paper, $\epsilon$ equals $0.2\hat{\mu}_{test,t}^m$ to provide sufficient sample points from the discrete probability bins to appropriately plot the reliability diagram and evaluate its diagonality.

## 4. Parameters for simulated PA data and BCNN

The simulations presented in *Section 2.2* in the main text used the parameters listed in Table S1 below, where key parameters for the simulated dataset and some training hyperparameters of BCNNs are specified.

Table S1. Parameters for simulated PA data and BCNN

| Type | Parameter | Value |
| --- | --- | --- |
| Raw data | Temporal samples | 2,048, |
| | Transducer element number | 128 |
| | Transducer central frequency | 15.625 MHz |
| | Temporal sampling rate | 62.5 MHz |
| | Transducer aperture size | 12.8 mm |
| | Transducer element pitch | 0.1 mm |
| | Signal-to-noise ratio (max signal / noise std) | 10-35 dB |
| PA image | Image number | 16,000 |
| | Signal dynamic range | 20 dB |
| | Mean power | 1 |
| | Axial samples (depth) | 512 (25.6 mm) |
| | Lateral samples (length) | 128 (12.7 mm) |
| | Vascular diameter | 0.05-0.3 mm |
| BCNN | Epoch number | 1,000 |
| | Early stop patience | 50 |
| | Learning rate | 0.0005 |
| | Batch size | 8 |
| | Trainable parameters (Hybrid-BCNN) | 7,803,081 |
| | Trainable parameters (Lap-BCNN) | 7,802,790 |

## 5. Comparison between Hybrid-BCNNs with Laplacian-distributed and Gaussian-distributed likelihood functions

In Hybrid-BCNN, segmentation and its uncertainty are the same for Gaussian-distributed and Laplacian-distributed likelihood functions. For PA image reconstruction and its uncertainty, as derived in a previous publication [5], the multivariate Gaussian-distributed likelihood function is:

$$p_{Gauss}(y_{image}|x, W_2) = \prod_{m=1}^{M} p_{Gauss}(y_{image}^m|x, W_2) \quad (S17)$$

$$p_{Gauss}(y_{image}^m|x, W_2) = \frac{1}{\sqrt{2\pi}\sigma^m} \exp\left(-\frac{(y_{image}^m - \mu_2^m)^2}{2(\sigma^m)^2}\right) \quad (S18)$$

where all variables are the same as in Eqs. (S4) and (S5) except that $\mu_2^m$ and $\sigma^m$ are the mean and standard deviation of the Gaussian for $y_{image}^m$.

Similar to Eqs. (S6-S8), the joint multivariate Bernoulli-distributed (for PA segmentation) and Gaussian-distributed (for PA image) likelihood function can be defined as,

$$p_{joint-Gauss} = p_{Ber}(y_{seg}|x, W_1) p_{Gauss}(y_{image}|x, W_2, y_{seg}) = \prod_{m=1}^{M} p_{Ber}(y_{seg}^m|x, W_1) p_{Gauss}(y_{image}^m|x, W_2, y_{seg}^m) \quad (S19)$$

$$p_{Gauss}(y_{image}^m|x, W_2, y_{seg}^m = 1) = \frac{1}{\sqrt{2\pi}\sigma^m} \exp\left(-\frac{(y_{image}^m - \mu_2^m)^2}{2(\sigma^m)^2}\right) \quad (S20)$$

$$p_{Gauss}(y_{image}^m|x, W_2, y_{seg}^m = 0) = \emptyset \quad (S21)$$

where all variables are the same as in Eqs. (S6-S8) except that $\mu_2^m$ and $\sigma^m$ are the mean and standard deviation of the Gaussian for $y_{image}^m$.

Similar to Eq. (S9), by taking logarithm and negative operations on Eq. (S19), the loss function $L_{Hybrid-Gauss}(W|x,y)$ is,

$$L_{Hybrid-Gauss}(W|x_n, y_n) = \sum_{m=1}^{M} \left[ (y_{seg,n}^m - 1)\log(1 - \mu_1^m) - y_{seg,n}^m \log(\mu_1^m) + y_{seg,n}^m \left( \frac{(y_{image,n}^m - \mu_2^m)^2}{2(\sigma^m)^2} + \log(\sqrt{2\pi}\sigma^m) \right) \right] \quad (S22)$$

where all variables are the same as in Eq. (S9) except that $\mu_2^m$ and $\sigma^m$ are the mean and standard deviation of the Gaussian for $y_{image,n}^m$.

Hybrid-BCNN with the Gaussian-distributed likelihood function also has three output channels, where one ($\mu_1$) is for the Bernoulli-distributed likelihood function and two ($\mu_2$ and $\sigma$) are for the Gaussian-distributed likelihood function. Its PA segmentation, PA image and predicted segmentation uncertainty are the same as Hybrid-BCNN with the Laplacian-distributed likelihood function in Eqs. (S10) and (S11).

The predicted uncertainty $\hat{\sigma}_{test,t(Gauss)}^m$ of the $m$th pixel for testing data $x_{test,t}$ for the Gaussian-distributed likelihood function is,

$$\hat{\sigma}_{test,t(Gauss)}^m = \sqrt{Var(y_{image}^m|x_{test,t}, X, Y)} = \sqrt{\mathbb{E}[Var(y_{image}^m|x_{test,t}, W)] + Var(\mathbb{E}[y_{image}^m|x_{test,t}, W])}$$

$$\approx \sqrt{\frac{1}{K}\sum_{k=1}^{K}(\hat{\sigma}_{test,t}^{m,k})^2 + \frac{1}{K}\sum_{k=1}^{K}(\hat{\mu}_{2test,t}^{m,k} - \hat{\mu}_{2test,t}^m)^2} \quad (S23)$$

where all variables are the same as in Eq. (S12) except that for the Gaussian-distributed likelihood function. $\hat{\sigma}_{test,t(Gauss)}^{m(D)} = \sqrt{\frac{1}{K}\sum_{k=1}^{K}(\hat{\sigma}_{test,t}^{m,k})^2}$ denotes data uncertainty, and $\hat{\sigma}_{test,t(Gauss)}^{m(M)} = \sqrt{\frac{1}{K}\sum_{k=1}^{K}(\hat{\mu}_{2test,t}^{m,k} - \hat{\mu}_{2test,t}^m)^2}$ denotes model uncertainty.

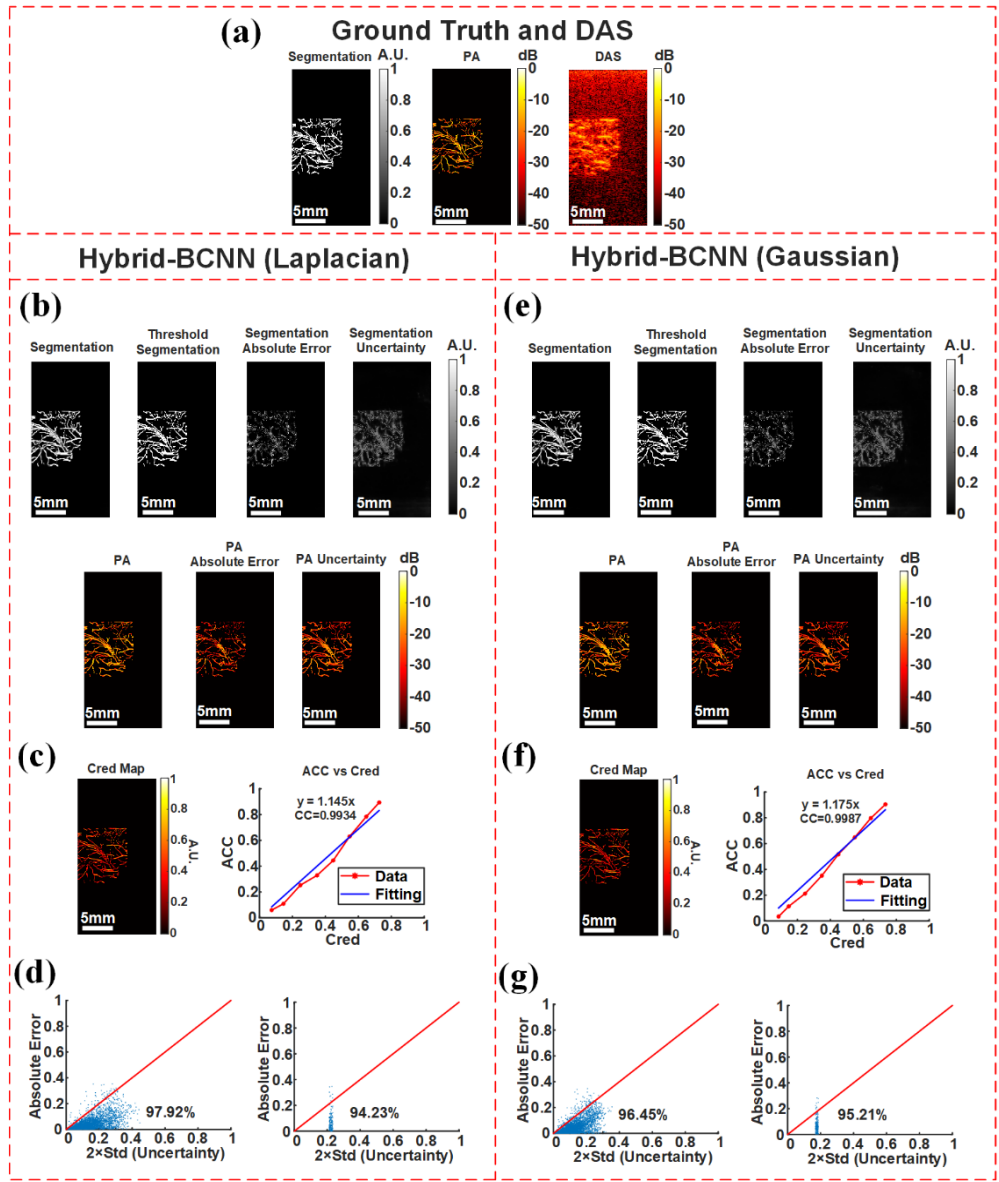

Fig. S1. Hybrid-BCNN (with Laplacian-distributed and Gaussian-distributed likelihood functions) results on simulated PA data. (a) A representative ground-truth PA segmentation and image, and the corresponding DAS-reconstructed PA image. (b) Results from Hybrid-BCNN with the Laplacian-distributed likelihood function. (c) Cred Map and ACC vs Cred for results in (b). (d) Absolute Error vs 2× predicted uncertainty for all pixels or pixels within a specific range of 2× predicted uncertainties [95% × ½ max(2×uncertainty), 105% × ½ max(2×uncertainty)] (within predicted segmentation regions) for results in (b). (e), (f) and (g) are the same as (b), (c) and (d) except that they are for Hybrid-BCNN with the Gaussian-distributed likelihood function. The scale bars denote 5 mm.

Figure S1 shows results for a representative sample in the testing dataset from Hybrid-BCNNs with Laplacian-distributed and Gaussian-distributed likelihood functions. They show that the two Hybrid-BCNNs both accurately produce the PA segmentation and image, and the corresponding uncertainties, with similar performance. The uncertainty predictions of PA segmentation and image for the two Hybrid-BCNNs are both highly statistically correlated to the segmentation/PA absolute errors. The corresponding credibility maps (Cred Maps) and reliability diagrams (ACC vs Cred) are shown in Fig. S1 (c) and (f), and the corresponding absolute error vs 2× predicted uncertainty are shown in Fig. S1 (d) and (g), where the two Hybrid-BCNNs still exhibit similar performance.

Quantitative metrics (segmentation accuracy, peak signal-to-noise ratio (PSNR), segmentation correlation coefficients (CC), CC and slopes of the ACC vs Cred reliability diagrams) were computed for all samples in the testing dataset to evaluate the two Hybrid-BCNNs, as shown in Table S2 (average value and standard deviation (in the parentheses) for each metric). For segmentation uncertainties, since the ground-truth only has two values (0 and 1), CC is sufficient to evaluate the uncertainty accuracy. For PA segmentation, the high segmentation accuracy and CC in Table S2 show that the two Hybrid-BCNNs both make accurate and similar reconstructions of the PA segmentation and its uncertainty. For PA image reconstruction and its uncertainty, the two Hybrid-BCNNs still show accurate and similar performance.

These comparisons show that the Hybrid-BCNN is robust to the specific probability distribution function for PA image reconstruction and its uncertainty as long as the selected function is reasonable. The Laplacian-distributed likelihood function is chosen for the results presented in the main text because of its slightly better performance in simulations.

**Table S2. Quantitative comparisons of Hybrid-BCNNs with Laplacian-distributed and Gaussian-distributed likelihood functions.**

**(PSNR (peak signal-to-noise ratio), CC (correlation coefficient), ACC vs Cred (ACC vs Cred reliability diagram))**

|  | Metrics | Hybrid-BCNN (Laplacian) | Hybrid-BCNN (Gaussian) |
|---|---|---|---|
| PA Segmentation and Image | Segmentation Accuracy | 0.9809 (0.0188) | 0.9805 (0.0191) |
|  | Image PSNR (dB) | 28.9855 (5.2125) | 28.7282 (5.1918) |
| Uncertainties | Segmentation CC | 0.7653 (0.0391) | 0.7419 (0.0534) |
|  | ACC vs Cred CC | 0.9901 (0.0213) | 0.9884 (0.0572) |
|  | ACC vs Cred Slope | 1.1665 (0.0954) | 1.2086 (0.0945) |